\documentclass[%
 reprint,
 twocolumn,
superscriptaddress,
 amsmath,amssymb,
prl,
]{revtex4-2}
\usepackage{epsfig,graphicx,amssymb,color,bm,dcolumn}
\usepackage{amsmath,amsbsy,amsopn, amstext,float,amsfonts,setspace,booktabs,url,latexsym,threeparttable}
\usepackage[colorlinks, linkcolor=blue, urlcolor=blue, citecolor=blue, unicode=true]{hyperref}

\newcommand{\eref}[1]{Eq.~(\ref{#1})}

\newcommand{\fref}[1]{Fig.~\ref{#1}}

\begin{document}
\title{Quantum control and signal enhancement exploiting the Stokes–anti-Stokes coherence} 
\author{Wen-Zhao Zhang}
\affiliation{Institute of Fundamental Physics and Quantum Technology, and Department of Physics, School of Physical Science and Technology, Ningbo University, Ningbo 315211, China}
\affiliation{Zhejiang Key Laboratory of Micro-Nano Quantum Chips and Quantum Control, and School of Physics, Zhejiang University, Hangzhou 310027, China}
\author{Keye Zhang}\thanks{kyzhang@phy.ecnu.edu.cn}
\affiliation{Quantum Institute for Light and Atoms, School of Physics, East China Normal University, Shanghai 200062, China}
\affiliation{Shanghai Branch, Hefei National Laboratory, Shanghai 201315, China}
\author{Jie Li}\thanks{jieli007@zju.edu.cn}
\affiliation{Zhejiang Key Laboratory of Micro-Nano Quantum Chips and Quantum Control, and School of Physics, Zhejiang University, Hangzhou 310027, China}
\begin{abstract}
We present a theoretical framework for the coherent coupling between Stokes and anti-Stokes scattering processes, revealing interference phenomena inaccessible to either process alone. Within a dispersive-interaction model beyond the resolved-sideband limit, we show that classical driving and system linewidth coherently links the two channels, enabling phase-controlled interference. 
Destructive interference induces intrinsic asymmetry in dispersively coupled systems, enabling coherent control of quantum information storage and transfer, while constructive interference leads to exponential signal amplification and thus enhanced quantum detection. 
This work establishes a unified picture for understanding Stokes–anti-Stokes coherence as a fundamental mechanism underlying both quantum control and metrology. Furthermore, it suggests that these functionalities can be further enhanced by implementing Stokes–anti-Stokes arrays.
\end{abstract}
\maketitle
\textit{Introduction}--Quantum coherence, a cornerstone of quantum mechanics rooted in the principle of state superposition, serves as a critical resource for various quantum technologies including quantum information processing \cite{PhysRevLett.110.120503,PhysRevLett.109.013603,RevModPhys.84.621,RevModPhys.84.777}, quantum computing \cite{nature.478.360,RevModPhys.79.135}, and precision measurement applications \cite{science.306.1330,RevModPhys.82.1155}. 
Quantum coherence can be manifested through the interaction between light and matter.
A representative platform for exploring these mechanisms is inelastic light–matter scattering \cite{RevModPhys.58.699}, where different scattering channels yield distinct coherent phenomena.
In particular, Stokes processes dominate energy release from light to matter, whereas anti-Stokes processes correspond to the energy absorption from matter to light.
The dichotomy between these interaction mechanisms highlights the rich physics emerging from different systems \cite{PhysRevLett.129.033202,PhysRevA.108.L051501,PhysRevResearch.3.L032010}, offering versatile tools for manipulating quantum states.

In recent years, dispersively coupled systems, such as optomechanical, magnomechanical, and optomagnonic systems, have emerged as a versatile platform for exploring analogous phenomena \cite{RevModPhys.86.1391,NJP.26.031201,APE.12.070101}. 
In these systems, where high-frequency excitations are scattered by low-frequency excitations, the effective interaction Hamiltonian can be two distinct types, yielding two sidebands: the red (Stokes) and blue (anti-Stokes) sidebands. 
The anti-Stokes process governs energy-exchange interactions and enables applications such as quantum state transfer \cite{PhysRevA.68.013808} and sideband cooling \cite{PhysRevLett.99.093901,PhysRevLett.99.093902}.
In contrast, the Stokes process dominates parametric amplification interactions and can generate entanglement and squeezing \cite{Nature.530.313}.
By appropriate protocol design, multiple interacting processes can be engineered to coexist within the system, thereby enhancing its quantum capabilities \cite{PhysRevLett.131.143603,PhysRevLett.108.120602,PhysRevLett.108.120602,PhysRevA.104.033705}. For example, introducing squeezing during detection can suppress quantum noise \cite{SCP.67.100313}, while beam-splitter–type interactions between multiple modes can achieve nonreciprocal behavior \cite{PhysRevA.96.013808,PhysRevA.91.053854}. 

\begin{figure}
  \centering
  \includegraphics[width=\linewidth]{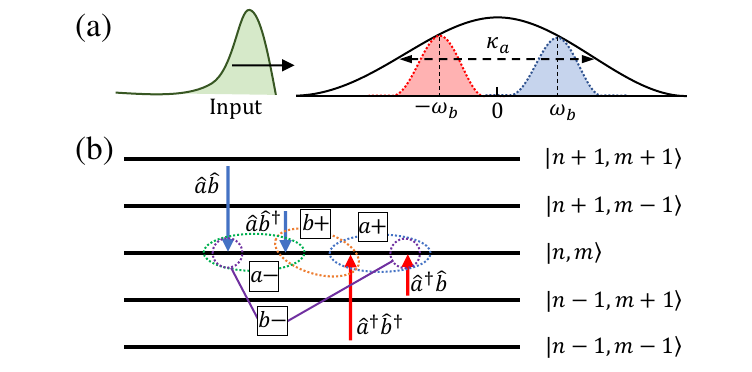}
  \caption{
  (a) Schematic diagram of the scattered Stokes and anti-Stokes sidebands in a dispersively coupled system under a classical driving field. 
  Two sidebands are within the mode profile in the unresolved-sideband limit.
  (b) Quantum interaction between the high- and low-frequency modes and the corresponding diagram of the energy level structure.
  }\label{fig1}
\end{figure}

To precisely control the scattering processes, such schemes typically require operations in the resolved-sideband regime \cite{Nphys.5.489,PhysRevLett.75.4011,PhysRevLett.99.093902}, where the sideband separation significantly exceeds the linewidth of the high-frequency mode, ensuring suppression of undesired scattering pathways.
Such requirements impose stringent experimental constraints, e.g., high quality factors of optical or microwave resonators \cite{RevModPhys.86.1391}, and preclude the dynamical advantages—such as rapid stabilization—that arise in strongly dissipative regimes \cite{PhysRevLett.110.153606}.
By contrast, in the unresolved-sideband regime the Stokes and anti-Stokes processes necessarily coexist.
These distinct physical effects—energy exchange and parametric amplification—can lead to the superposition of dynamically generated quantum states, thereby introducing nontrivial quantum coherence between these two processes. 
This is illustrated in \fref{fig1}a, where the linewidth of the high-frequency mode $a$ spans two sidebands scattered by the low-frequency mode $b$, simultaneously enhancing both Stokes and anti-Stokes processes.
Under coherent driving, this mechanism opens four excitation channels through dispersive coupling, which are illustrated in the corresponding energy-level diagram (\fref{fig1}b).
In the double-excitation subspace, the number states of modes $a$ and $b$ are denoted as $|n\rangle$ and $|m\rangle$, respectively.
The corresponding interaction Hamiltonian then contains four distinct transitions.
As indicated by the red arrows, both rotational-wave and counter-rotational-wave processes increase the excitation number of mode $a$, driving the transition $|n-1\rangle \!\to\! |n\rangle$.
The rotational-wave process transfers an excitation from mode $b$ to mode $a$, whereas the counter-rotational-wave process, under a classical drive, simultaneously creates a pair of excitations in modes $a$ and $b$.
These two processes allow different lower-lying excitations of mode $a$ to coherently merge into the common final state $|n,m\rangle$, thereby establishing multipath quantum coherence (labeled `$a+$' in \fref{fig1}b).
A symmetric process corresponding to the annihilation of an excitation in mode $a$ is indicated by the blue arrows and gives rise to the multipath coherence labeled `$a-$'.
Similarly, the multipath coherences associated with the creation and annihilation of an excitation in mode $b$ are labeled `$b+$' and `$b-$', respectively.
Such coherence between multiple physical processes behaves differently in the resolved- and unresolved-sideband regimes.
In the former, the linewidth of the high-frequency mode $a$ is much smaller than the sideband separation, so a given drive excites only one scattering channel, cutting off the interference pathways in $a\pm$ and $b\pm$. 
By contrast, in the latter, the linewidth spans both sidebands, allowing the Stokes and anti-Stokes scattering channels simultaneously, activating all four multipath coherences.
The mechanism underlying this hybrid coherence remains elusive.
In this Letter, we explore Stokes–anti-Stokes coherence (SASC) in a widely used dispersive-coupling model, applicable to systems such as optomechanics \cite{RevModPhys.86.1391} and magnomechanics \cite{NJP.26.031201}. 
We show that SASC can be harnessed to enhance, suppress, or coherently control physical processes, particularly in transmission, detection, and related tasks, and can be extended to array structures for further enhanced performance.

\textit{Dynamical characteristics of dispersively coupled units}--We first consider the minimal model: a dispersively coupled unit (DU) under the weak-coupling and strong driving condition. 
The corresponding Hamiltonian reads
\begin{equation}\label{eq1}
  \frac{\hat{H}}{\hbar}=\sum_{j=a,b}\omega_{j}\hat{j}^{\dag}\hat{j}
  +g_{ab}\hat{a}^{\dag}\hat{a}(\hat{b}+\hat{b}^{\dag})+i(\varepsilon_a e^{-i \omega_d t} \hat{a}^{\dag} -\text{h.c.}),
\end{equation}
where mode $a$ is a high-frequency mode with frequency $\omega_a$, which can be an optical cavity mode \cite{RevModPhys.86.1391} or a magnon mode \cite{NJP.26.031201}, while mode $b$ is a relatively low-frequency mode with frequency $\omega_b$, which may correspond to movable mirrors \cite{Nature.444.71}, magnetostrictively induced vibration \cite{PhysRevLett.124.213604}, acoustic phonons confined by Bragg reflectors \cite{PhysRevLett.110.037403}, or low-frequency waveguides coupled via evanescent fields \cite{PhysRevLett.129.053901}.
The two modes are dispersively coupled with strength $g_{ab}$.
When mode $a$ is driven by an external drive with the frequency $\omega_d$ and the corresponding coupling strength $\varepsilon_a$, it interacts with mode $b$ through both types of interactions in the unresolved-sideband limit.
In this regime, the rotating wave approximation (RWA) breaks down. Nevertheless, the linearization treatment remains valid \cite{PhysRevLett.98.030405}, by expressing
$\hat{j} \rightarrow \langle \hat{j} \rangle + \delta \hat{j}$,
where $\langle \hat{j} \rangle$ denotes the mean value of mode $j$, and $\delta \hat{j}$ captures its quantum fluctuation. 
In what follows, we focus on quantum effects arising from these fluctuations, and obtain the corresponding linearized Langevin equations
\begin{eqnarray}\label{eq2}
\delta\dot{\hat{a}} &=& -(i\tilde{\Delta}_a+\frac{\kappa_a}{2})\delta\hat{a}-i G_{ab}(\delta\hat{b}^{\dag}+\delta\hat{b})+\sqrt{\kappa_a} \hat{a}_{\text{in}},\\
\delta\dot{\hat{b}} &=& -(i\omega_b+\frac{\kappa_b}{2})\delta\hat{b}-i (G_{ab}\delta\hat{a}^{\dag}+G_{ab}^{*}\delta\hat{a})+\sqrt{\kappa_b} \hat{b}_{\text{in}},\nonumber
\end{eqnarray}
where $\tilde{\Delta}_a=\omega_c-\omega_{d}+g_{ab}(\langle\hat{b}\rangle+\langle\hat{b}^{\dag}\rangle)$ is the effective detuning, $G_{ab} = g_{ab} \langle \hat{a} \rangle\equiv |G_{ab}|e^{i \theta}$ is the linearized complex coupling coefficient  \cite{Nature.568.65,nature.460.724}, with $\theta$ serving as a key parameter governing the interference between distinct scattering channels, which can be tuned continuously via the classical drive \cite{PhysRevA.110.043721}.
Equation~(\ref{eq2}) indicates that the phase response in the $b$-to-$a$ reaction fundamentally differs from that in the $a$-to-$b$.
This asymmetry is vital for the following discussion of coherent interactions.
Here $\kappa_j$ and $\hat{j}_{\text{in}}$ denote the damping rate and input noise operator of mode $j$, respectively.
By applying the input–output formalism, the vector of output operators $\hat{Q}_{\text{\text{out}}}=(\hat{a}_{\text{\text{out}}},\hat{a}_{\text{\text{out}}}^{\dag},\hat{b}_{\text{\text{out}}},\hat{b}_{\text{\text{out}}}^{\dag})^{\text{T}}$ in the frequency domain is obtained as
\begin{equation}
    \hat{Q}_{\text{out}}(\omega)=\Gamma(\omega) \hat{Q}_{\text{in}}(\omega),
\end{equation}
where $\hat{Q}_{\text{in}}=(\hat{a}_{\text{in}},\hat{a}_{\text{in}}^{\dag},\hat{b}_{\text{in}},\hat{b}_{\text{in}}^{\dag})^{\text{T}}$  is the input vector and the transformation matrix is given by $\Gamma(\omega)=L(i\omega \Lambda-M)^{-1}L-I$.
Here, $M$ is the drift matrix of \eref{eq2}, the damping matrix is $L=\text{diag} (\sqrt{\kappa_a},\sqrt{\kappa_a},\sqrt{\kappa_b},\sqrt{\kappa_b})$ and $\Lambda=\text{diag}(-1,1,-1,1)$.
 The input noise is typically Gaussian and satisfies $\langle \hat{j}_{\text{in}}^{\dag}(\omega)\hat{j}_{\text{in}}(\omega')\rangle=\delta(\omega+\omega')j_{\text{th}}$, $\langle \hat{j}_{\text{in}}(\omega)\hat{j}_{\text{in}}^{\dag}(\omega')\rangle=\delta(\omega+\omega')(j_{\text{th}}+1)$ and $\langle \hat{j}_{\text{in}}^{\dag}(\omega)\hat{j}_{\text{in}}^{\dag}(\omega')\rangle=\langle \hat{j}_{\text{in}}(\omega)\hat{j}_{\text{in}}(\omega')\rangle=0$. 
Here $j_{\text{th}} = 1/\left[\exp\left(\frac{\hbar \omega_j}{k_{\mathrm{B}} T}\right) - 1\right]$ denotes the thermal occupation with environment temperature $T$ and Boltzmann's constant $k_{\text{B}}$.
The fluctuation of the quadrature of the output field, $\hat{x}_j=(\hat{j}_{\text{out}}^{\dag} e^{i \psi}+\hat{j}_{\text{out}}e^{-i \psi})/\sqrt{2}$, can be measured through the Homodyne detection (HD), where $\psi$ represents the phase angle and can be adjusted by the local oscillator.
For convenience of discussion, we set $\psi=0$ here.
The corresponding output spectrum can be expressed as \cite{SMAppendix}:
\begin{eqnarray}
    S_{\text{out}}^j(\omega) &=& T_j S_{jj}(\omega)+\frac{T_{k+}+T_{k-}}{2} S_{kk}(\omega),
\end{eqnarray}
where $j,k \in \{ a,b \}$, $j\neq k$ and
$S_{jj}(\omega)=\left[\int d\omega'\langle \hat{j}^{\dag}_{\text{\text{in}}}(\omega')\hat{j}_{\text{\text{in}}}(\omega)\rangle+h.c.\right ]/2$ is the general input spectrum of mode $j$.
In optomechanical platforms, low-frequency phonons can be extracted directly from the Bragg reflector of a GaAs/AlAs microcavity \cite{PhysRevLett.110.037403}, or indirectly via circuit design \cite{nphys.9.480,Nature.464.697}.
The output spectra are characterized by the self-transformation coefficients $T_a = |\Gamma_{1,1}(\theta) + \Gamma_{2,1}(\theta)|^2$ and $T_b = |\Gamma_{3,3}(\theta) + \Gamma_{4,3}(\theta)|^2$, and by the intermode terms $T_{a+} = |\Gamma_{1,3}(\theta) + \Gamma_{2,3}(\theta)|^2$ and $T_{b+} = |\Gamma_{3,1}(\theta) + \Gamma_{4,1}(\theta)|^2$, which encode the coherent mixing of the Stokes and anti-Stokes processes. 
Here, $\Gamma_{i,j}(\theta)$ is the $(i,j)$-th element of $\Gamma(\omega)$, quantifying the contribution of the $j$-th operator in $\hat{Q}_{\text{in}}$ to the $i$-th operator in $\hat{Q}_{\text{out}}$.
While Hermitian symmetry ensures $T_{a+} = T_{a-}$ and $T_{b+} = T_{b-}$.
As illustrated in \fref{fig1}b, the phase $\theta$ enters differently in the two processes, contributing as a global phase in the ‘$a+$’ process but a local phase in the ‘$b-$’ process, yielding $T_{a+} \neq T_{b-}$.
This intrinsic phase asymmetry produces direction-dependent spectral responses that are unattainable in the conventional resolved-sideband regime \cite{PhysRevLett.132.210402}.
We quantify this effect using \cite{EPL.106.54003}
\begin{equation}\label{eq6}
    \mathcal{R}_{ab}=\frac{T_{a+}-T_{b- }}{T_{a+}+T_{b-}},
\end{equation}
where $\mathcal{R}_{ab}=0$ indicates symmetry, $\mathcal{R}_{ab}\neq 0$ signals SASC, and $\mathcal{R}_{ab} =\pm 1$ indicates that the processes labeled `${a\pm}$' or `${b\pm}$' in \fref{fig1}b are fully suppressed due to the complete destructive interference.
The phase dependence of $\mathcal{R}_{ab}$ away from zero provides a natural measure of the SASC strength.
\begin{figure}
  \centering
  \includegraphics[width=1\linewidth]{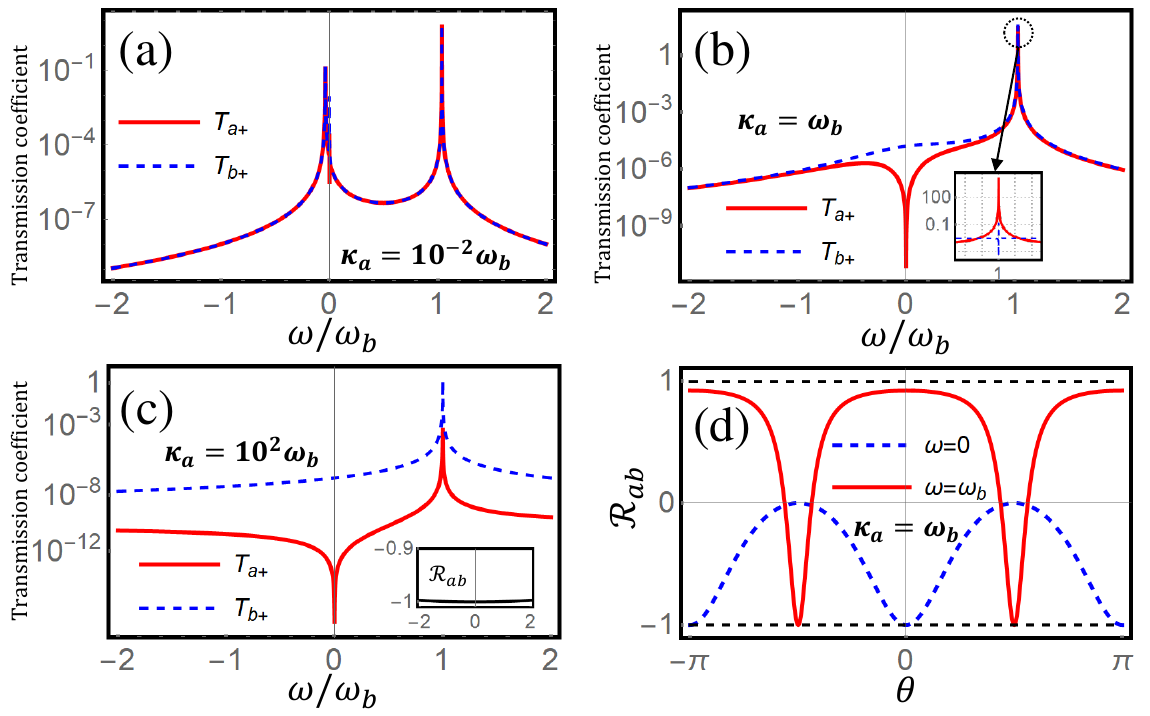}
  \caption{Transmission coefficients $T_{a+}$ and $T_{b+}$ for different dissipation ratios $\kappa_a / \omega_b$: 
(a) $10^{-2}$, (b) $1$ and (c) $10^2$.  
The inset in (c) shows the asymmetry factor $\mathcal{R}_{ab}$ for $\kappa_a=10^{2}\omega_b$.
(d) $\mathcal{R}_{ab}$ versus $\theta$ for $\kappa_a=\omega_b$ at $\omega=0$ and $\omega=\omega_b$.
Other parameters are $\Delta_a = 0$, $G_{ab} = 10^{-1}\omega_b$, $\kappa_b = 10^{-4}\omega_b$, $T = 10\,\mathrm{mK}$, and $\omega_b = 2\pi \times 10\,\mathrm{MHz}$.
We set $\psi=0$ in the figure. The effect of the local oscillation phase $\psi$ is analyzed in \cite{SMAppendix}.
}\label{fig2}
\end{figure}

\fref{fig2}a-d show the effect of the linewidth on the SASC-induced asymmetry. 
For $\kappa_a \ll \omega_b$, both sidebands are essentially suppressed, preventing the SASC thus yielding a symmetric response (\fref{fig2}a). 
As $\kappa_a$ increases to be comparable with $\omega_b$, which enables partial interference between the two scatterings, the asymmetry appears in a narrow spectral range (\fref{fig2}b). 
For a sufficient large $\kappa_a \gg \omega_b$, the interference becomes fully effective, yielding pronounced asymmetry over a broad range (\fref{fig2}c).
The corresponding behaviour of $\mathcal{R}_{ab}$ is shown in \fref{fig2}d. 
For $\kappa_a = \omega_b$, multipath coherence exhibits strong phase dependence, yielding a phase-dependent asymmetric factor. 

\textit{Applications of extended dispersively coupled units}--The SASC effect in a DU enables phase-modulated coherent quantum control. 
Extending a single unit to coupled DUs yields richer physics, and, as a demonstration, we reveal this in an opto-magnomechanical system \cite{NJP.26.031201,LRP.17.2200866}, which consists of three modes and both the magnon ($\hat{m}, \omega_m$) and cavity ($\hat{c}, \omega_c$) modes couple dispersively to a common mechanical resonator ($\hat{b}, \omega_b$). The interaction $g_{jb}(\hat{b}+\hat{b}^\dagger)\hat{j}^\dagger\hat{j}$ ($j=m,c$) is typically weak but can be enhanced through two external driving at $\omega_{dj}$, yielding effective couplings $G_j = |G_j|e^{i\theta_j}$.
 
The linearized Heisenberg–Langevin equations read $\dot{\hat{Q}}^{(3)} = M^{(3)} \hat{Q}^{(3)} + L^{(3)} \hat{Q}_{\mathrm{\text{in}}}^{(3)}$, with $\hat{Q}^{(3)} = (\delta \hat{m}, \delta \hat{m}^\dag, \delta \hat{b}, \delta \hat{b}^\dag, \delta \hat{c}, \delta \hat{c}^\dag)^{\mathrm{T}}$, $L^{(3)} = \mathrm{diag}(\sqrt{\kappa_m}, \sqrt{\kappa_m}, \sqrt{\kappa_b}, \sqrt{\kappa_b}, \sqrt{\kappa_c}, \sqrt{\kappa_c})$ and $M^{(3)}$ being the coefficient matrix \cite{SMAppendix}.
Using the input–output relation, we obtain $\hat{Q}_{\mathrm{\text{out}}}^{(3)} = \Gamma^{(3)}(\omega) \hat{Q}_{\mathrm{\text{in}}}^{(3)}(\omega)$ with $\Gamma^{(3)}(\omega) = L^{(3)} (i\omega \Lambda^{(3)} - M^{(3)})^{-1} L^{(3)} - I^{(3)}$, where $\Lambda^{(3)}=\text{diag}(-1,1,-1,1,-1,)$. 
We define the asymmetry parameter $\mathcal{R}_{ij} = (T_{i+} - T_{-j})/(T_{i+} + T_{-j})$, where $\mathcal{R}_{ij} = \pm 1$ corresponds unidirectional transmission.
For the three-mode system ordered as ($m,b,c$), interactions with left (right) neighbors are labeled by `$\pm j$' (`$j\pm $').
The relevant characteristic coefficients are $T_{m\pm} = |\Gamma_{1,3}^{(3)}(\theta_m) + \Gamma_{2,3}^{(3)}(\theta_m) |^2$, $T_{\pm b} = |\Gamma_{3,1}^{(3)}(\theta_m) + \Gamma_{4,1}^{(3)}(\theta_m) |^2$, $T_{b\pm} = |\Gamma_{3,5}^{(3)}(\theta_c)  + \Gamma_{4,5}^{(3)}(\theta_c) |^2$, and $T_{\pm c} = |\Gamma_{5,3}^{(3)}(\theta_c)  + \Gamma_{6,3}^{(3)}(\theta_c) |^2$.

\begin{figure}
    \centering
    \includegraphics[width=1\linewidth]{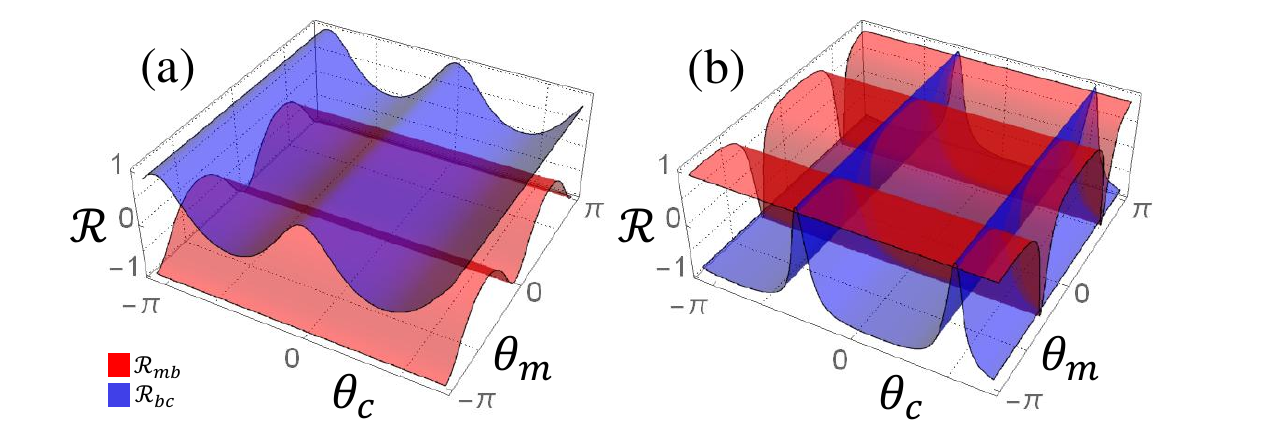}
    \caption{Asymmetric factors $\mathcal{R}_{mb}$ (red) and $\mathcal{R}_{bc}$ (blue) as functions of the coupling phases $\theta_m$ and $\theta_c$ for (a) $\omega=0$ and (b) $\omega=\omega_b$. 
   Other parameters are $\psi=0$, $\Delta_m=\Delta_c=0$, $G_c=G_m=10^{-1} \omega_b$, $\kappa_c=\kappa_m=\omega_b$, $\kappa_b=10^{-4}\omega_b$, $T=10$~mK, $\omega_b=2 \pi \times 10$~MHz and $\omega_m=2 \pi \times 10$~GHz.}
    \label{fig3}
\end{figure}

As shown in \fref{fig3}, the factor $\mathcal{R}_{mb}$ ($\mathcal{R}_{bc}$) is a periodic function of  $\theta_m$ ($\theta_c$), and $\mathcal{R}_{mb}$ and $\mathcal{R}_{bc}$ are independent with each other.
This is because the $m$–$b$ and $b$–$c$ interactions are fully governed by $G_m$ and $G_c$, of which the phases can be tuned separately. 
The system can thus be switched between an $m \! \to \!  b \!\to \!c$ configuration with $\mathcal{R}_{mb}=\mathcal{R}_{bc}=-1$, enabling magnon-to-optical transmission, and an $m \!\to\!  b\! \leftarrow \! c$ configuration with $\mathcal{R}_{mb}=-\mathcal{R}_{bc}=-1$, suitable for quantum storage by exploiting the long lifetime of the mechanical mode \cite{ncomm.13.1507}.
This behavior is consistent with that shown in \fref{fig2}d. 
For $\omega=0$ (\fref{fig3}a), $\mathcal{R}$ cannot be switched between $1$ and $-1$, realizing only the function of quantum storage. 
In contrast, for $\omega=\omega_b$ (\fref{fig3}b), full switching is achieved, enabling the application, e.g., optical readout of the solid-state magnonic states \cite{QST.8.015014}.

Figure~\ref{fig3} demonstrates that, by modulating the two phases, the SASC-induced coherent cancellation can be exploited to manipulate the interaction directionality and realize different functions, on demand, in the same system.
In contrast, when the interference becomes constructive, the SASC can lead to coherent enhancement, which we show below can be exploited to enhance the signal-to-noise ratio (SNR) in related detection applications.
This enhancement can be expanded between DUs, allowing constructive interference across units.
Since this enhancement of SASC is cumulative, we present the results only for a dual-DU system and discuss in the end the results of an extended chain structure. 
Using the same opto-magnomechanical (dual-DU) model for signal detection, we consider, without loss of generality, a weak (magnetic) signal is applied onto the magnon mode, and the signal is detected in the optical output field.
Combined with balanced HD, the amplification and SNR spectra for a {\it unit} input signal are given by \cite{SMAppendix},
\begin{eqnarray}\label{eqsadd}
    S_{\text{AP}}(\omega) &=& |C_s(\omega)+C_s^*(\omega)|^2,\\
    S_{\text{SNR}}(\omega)&=&\left\{\frac{1}{2}\sum_{j=c,m,b} \left| \frac{C_j(\omega)}{C_s(\omega)+C_s^*(\omega)} \right|^2 (2j_{\text{th}}+1)\right\}^{-1}, \nonumber
\end{eqnarray}
where $C_j(\omega)$ is the coefficient of $\hat{j}_{\text{in}}~(j=c,m,b,s)$ appearing in the quadrature of the output field.
The system input operator $\hat{m}_{\text{in}}$ consists of two parts: a signal input $\hat{s}_{\text{in}}$ carrying quantum information, and a noise input $\hat{m}_{\text{th}}$. 
The signal input $\hat{s}_{\text{in}}$ is treated as a real observable satisfying $\hat{s}_{\text{in}}^{\dag} = \hat{s}_{\text{in}}$ and $\langle \hat{s}_{\text{in}}^{\dag}\hat{s}_{\text{in}}\rangle=s$ \cite{PhysRevA.109.023709,PhysRevA.102.023525,PhysRevA.99.063811} (see \cite{SMAppendix} for the results of a complex signal by including the phase factor).
We consider a cryogenic environment at $T= 10$~mK, where the optical and magnonic modes are essentially in vacuum state, while the mechanical mode is in a thermal state.
The amplification coefficient $S_{\text{AP}}(\omega) \propto|(G_m^* \chi_m + G_m \chi_m^{\dag})(G_c \chi_c + G_c^* \chi_c^{\dag})|^2$ shows that the signal is jointly shaped by magnonic and optical multipath coherences, with $\chi_j=\left[ i(\Delta_j-\omega)+\kappa_j/2 \right]^{-1}$ and  $\Delta_j=\omega_j-\omega_{dj}+g_{jb}(\langle\hat{b}\rangle+\langle\hat{b}^{\dag}\rangle),~\{j=m,c\}$.
Note that this coherently enhances the amplification coefficient rather than suppressing the noise, still yielding an enhanced SNR due to the uncorrelated nature of the noise. 
By selecting an appropriate phase $\theta_j$, perfect constructive interference is achieved, yielding a SASC-enhanced amplification relative to the resolved-sideband situation \cite{PhysRevA.90.043848,PhysRevA.99.063811}.
\begin{figure}[tb]
    \centering
    \includegraphics[width=1\linewidth]{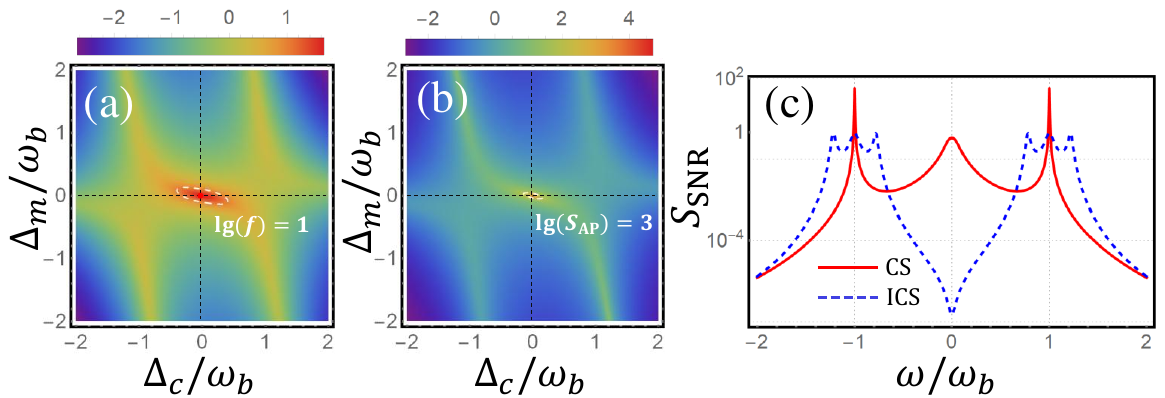}
    \caption{(a) Density plot of $\lg(f)$ as a function of $\Delta_m$ and $\Delta_c$, where larger values of $f$ correspond to amplified signal in the CS scheme. 
(b) Logarithmic amplification factor $\lg(S_{\text{AP}})$ as a function of $\Delta_m$ and $\Delta_c$.
(c) Comparison of the SNR spectra for the ICS and CS schemes at $\Delta_m = \Delta_c = 0$.
Other parameters: $G_c =10^{-1}\omega_b$, $G_m =2\times 10^{-1}\omega_b $, $\kappa_c = 10^{-1}\omega_b$, $\kappa_m = \omega_b$, $\kappa_b = 10^{-4}\omega_b$, $T = 10~\mathrm{mK}$, $\omega_b = 2\pi \times 10~\mathrm{MHz}$ and $\omega_m = 2\pi \times 10~\mathrm{GHz}$.}
    \label{fig4}
\end{figure}

An intuitive comparison is presented in \fref{fig4}, where we contrast our coherent-scattering (CS) scheme with the conventional incoherent-scattering (ICS) approach operating in the resolved-sideband regime \cite{PhysRevLett.115.211104,PhysRevApplied.17.034020}.
The definition of \eref{eqsadd} enables us to compare different protocols. 
We introduce a proportionality factor $f = S_{\text{SNR}}^{\text{CS}}/S_{\text{SNR}}^{\text{ICS}}$,
where $S_{\text{SNR}}^{\text{ICS}}$ denotes the maximum SNR under the ICS scheme with $\kappa_c=\kappa_m=0.1 \omega_b,~\Delta_c=\Delta_m=\omega_b$ (other parameters are in Fig.~\ref{fig4}), and $S_{\text{SNR}}^{\text{CS}} = \max[S_{\text{SNR}}(\omega)]$ represents the maximum SNR for given parameters $(\Delta_c,\Delta_m)$.
Since detection is performed at the optical output port, using a small decay rate $\kappa_c = 0.1 \omega_b$ yields the optimal tradeoff between signal amplification and noise increase.
As shown in \fref{fig4}a, a large parameter regime exhibits $f > 1$ (orange–red areas), indicating the advantage of the CS scheme.
The optimal condition occurs around $(\Delta_c,\Delta_m) = (0,0)$, where both the magnonic and optical modes are resonant with the drives.
The corresponding amplification contour plot in \fref{fig4}b coincides with the high-$f$ regions in \fref{fig4}a, revealing that the SNR enhancement stems from signal amplification rather than coherent noise cancellation, consistent with our earlier analysis.
Furthermore, as shown in \fref{fig4}c, our protocol achieves substantial SNR enhancement and significantly outperforms the ICS scheme.

\textit{Discussion and conclusion}--The extension of DUs, such as a chain structure, can achieve an enhanced performance of the functionalities. 
Consistent with the behavior observed in \fref{fig3} and \fref{fig4}, each unit can be controlled independently by its local drive. 
This facilitates key operations such as transmission regulation and signal amplification.
While the signal gain scales exponentially with the number of DUs $N$, the magnitude of this scaling depends heavily on the driving scheme. 
By alternating the driving detunings, the system acts as a simple tandem arrangement of amplification and transmission units, with the gain scaling as $\mathcal{A}^N$. 
For the parameters in \fref{fig4}, this yields a relatively limited base of $\mathcal{A} \approx 3.68$ \cite{SMAppendix}. 
In sharp contrast, a far more potent amplification is achieved via amplitude-modulated drives that break time-reversal symmetry. 
This enables a more stringent global matching condition, $\left| \chi_j \chi_q [G^*_{j,j}(-\omega)G_{j,j+1}(-\omega) - \text{c.c.}] \right| \equiv \mathcal{G}(\omega)$ for all modes (where $G_{i,j}$ is the linearized coupling), allowing the gain to scale as $\mathcal{G}^N$. 
For an optomechanical array at the mechanical sideband, the same driving fields ($1064$~nm, $10$~mW) yield $\mathcal{G} \approx 450$ \cite{SMAppendix}, a base two orders of magnitude larger than the former case.

In conclusion, the DU structure exhibits clear SASC effects in the unresolved-sideband regime, enabling phase-controlled interference to be tuned for either cancellation or enhancement as needed. The resulting capabilities including direction-steerable transmission and signal amplification require no strict sideband control, high-$Q$ resonators, or precise quantum state engineering, significantly lowering experimental barriers. 
Moreover, the independent tunability of the interference phase via classical drives ensures excellent scalability, positioning this system as a promising building block for hybrid quantum networks  \cite{Nature.484.195}. 
Analysis of the intrinsic SASC identifies the coupling phase and system linewidth as the core factors governing the coherence. 
These insights may deepen our understanding of quantum interference in systems where conjugate processes coexist and interact, enabling effective quantum control and signal manipulation in practical devices.

\textit{Acknowledgments}--This work was supported by the National Natural Science Foundation of China (Grant Nos.~12474365, 12374328 and 12234014), the Zhejiang Provincial Natural Science Foundation of China (Grant No.~LR25A050001), the Quantum Science and Technology–National Science and Technology Major Project (No.~2021ZD0303200) and the
Science and Technology Innovation Plan of Shanghai Science and Technology Commission (No.~24LZ1400600).

\bibliography{ref}

\end{document}